\documentclass[11pt,letterpaper]{paper}
\usepackage[margin=1.25in]{geometry}
\usepackage{authblk}
\usepackage{graphicx,xcolor,empheq}
\usepackage{amssymb,amsmath,amsthm,bm}
\usepackage[sorting=none,style=numeric-comp]{biblatex}
\usepackage{hyperref}

\begin{document}

\title{The Mechanism of Shrimpoluminescence}

\author{\raggedright Tyler C. Sterling  \thanks{ty.sterling@colorado.edu}}
%\author{Tyler C. Sterling}
%\email{ty.sterling@colorado.edu}
\affil{\raggedright Department of Physics, University of Colorado Boulder, Boulder CO, 80309, USA}

\maketitle
{\raggedright \mdseries \upshape \today}

\vspace{0.5cm} \noindent
\emph{Preface: This brief review paper was submitted to the physics department at the University of Colorado Feb. 2022 to satisfy qualifying exam requirements during my PhD.}

\begin{abstract}
Snapping shrimp produce bubbles that emit light when they collapse. When a bubble collapses so strongly that it emits light, the light emission is usually called sonoluminescence; in the case of the shrimp, it is called \emph{shrimpoluminescence}. The bubble collapses so fast that no heat can escape and the gas trapped in the bubble becomes hot enough to ionize. Light is emitted through electron-ion bremsstrahlung, electron-atom bremsstrahlung, and electron-ion recombination. In this paper, we study the dynamics of a sonoluminescing bubble and learn how to calculate the spectrum of emitted light, allowing us to explain the physical mechanisms of shrimpoluminescence.
\end{abstract}

\begin{figure}
\centering
\includegraphics[width=1\linewidth]{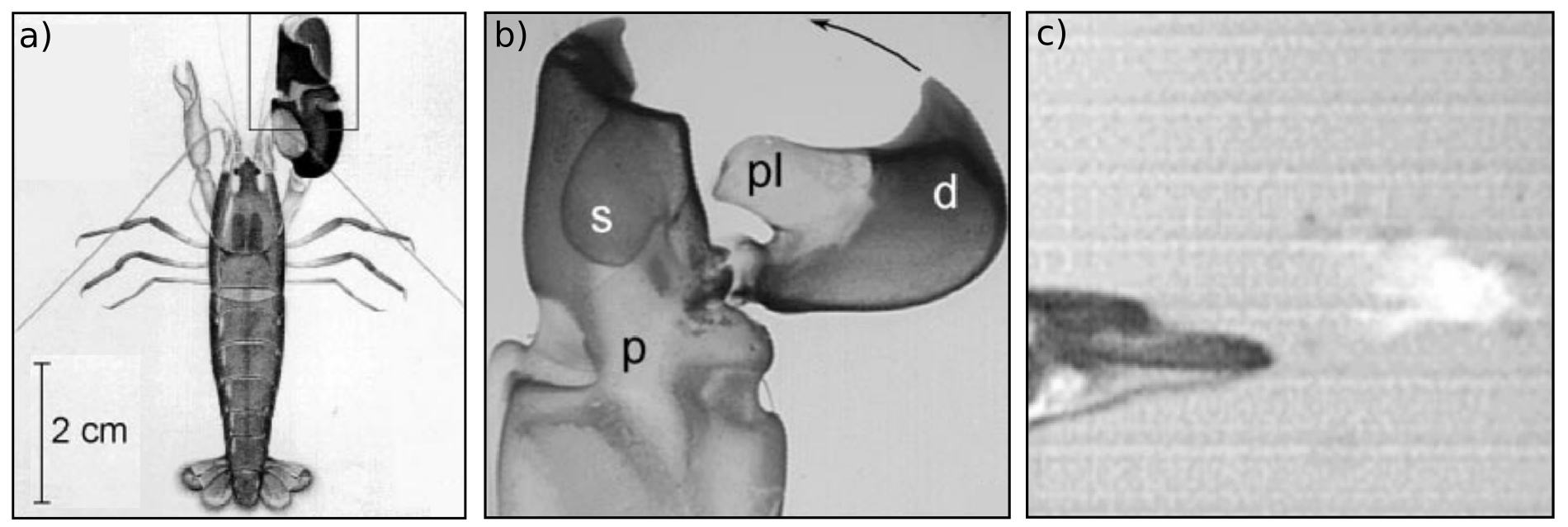}
    \caption{(left) Snapping shrimp (\emph{Alpheus heterochaelis}). (center) Blown-up view of the shrimp's claw. The \emph{plunger} (pl) on the \emph{dactyl} (d) rapidly enters the \emph{socket} (s), ejecting a high-velocity jet of water. A time near the bubble collapse is shown on the right. The light emission is too dim to be seen by the naked eye. Adapted from refs. \cite{versluis2000snapping} and \cite{lohse2001snapping}.}
\label{fig:shrimp}
\end{figure}

\section{Introduction}

Snapping shrimp, like our cute friend in Fig. \ref{fig:shrimp}, produce cavitating bubbles by snapping their claws \cite{versluis2000snapping,lohse2001snapping,tang2019bioinspired}. They have a strong appendage called the \emph{dactyl} [Fig. \ref{fig:shrimp} (center)] that is used to create a high-velocity jet of water. The low pressure region in the jet's wake forms a bubble [Fig. \ref{fig:shrimp} (right)] that, when it collapses, produces a noise loud enough to be detected over a mile away \cite{everest1948acoustical}. The sound wave produced by the collapsing bubble is used to stun or kill prey \cite{versluis2000snapping}. If the shrimp's prey had very sensitive eyes (and also were not dead) they might notice a flash of light is also produced through an effect known as ``shrimpoluminescence" in the case of the shrimp \cite{lohse2001snapping}, but more generally called \emph{sonoluminescence}.
%The noise produced by groups of shrimp is so intense that the U.S. Navy used them as ``sonar-camouflage" in the Pacific ocean during World War II \cite{versluis2000snapping}. The shrimp were not patriots helping the war-effort however; they snapped for food. The sound wave produced by the cavitating bubble is used to stun or kill prey \cite{versluis2000snapping}. If the shrimp's prey had very sensitive eyes (and also were not dead) they might notice a flash of light is also produced through an effect referred to as ``shrimpoluminescence" in the case of the shrimp \cite{lohse2001snapping}, but more generally known as \emph{sonoluminescence}.

\begin{figure}
\centering
\includegraphics[width=0.6\linewidth]{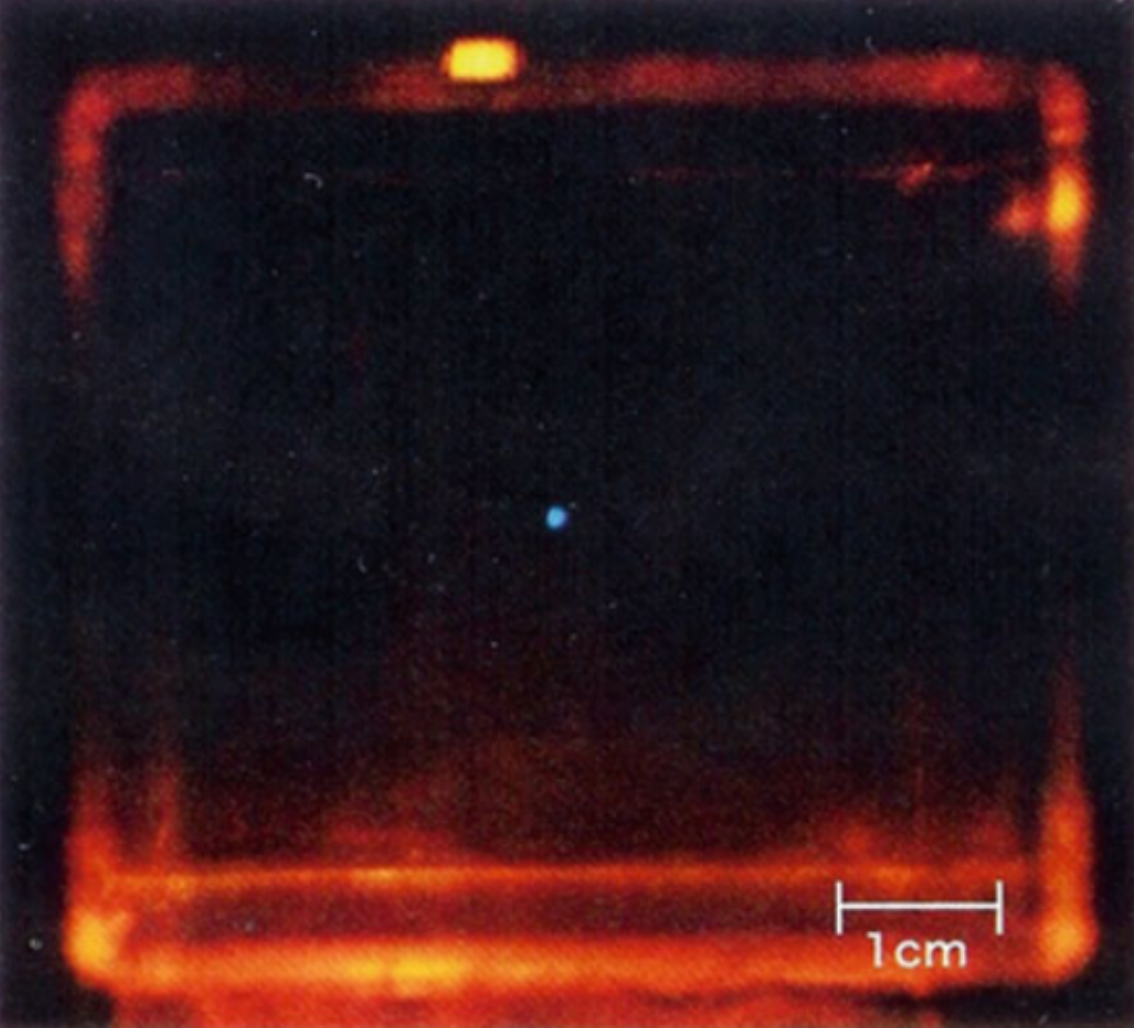}
    \caption{A single stable sonoluminescing bubble trapped in a flask containing water. Adapted from ref. \cite{yasui2018acoustic}.}
    \label{fig:sbsl}
\end{figure}

Sonoluminescence (SL) is defined as the process by which a ``driven gas bubble collapses so strongly that the energy focusing at collapse leads to light emission" \cite{brenner2002single}. SL comes in two forms: (i) single-bubble sonoluminescence and (ii) multi-bubble sonoluminescence. Multi-bubble sonoluminescence (MBSL) consists of  ``the simultaneous creation and destruction of many separate, individual cavitation bubbles" \cite{crum1994sonoluminescence,brenner2002single}. In single-bubble sonoluminescence (SBSL), rather obviously, only a single bubble is present \cite{gaitan1992sonoluminescence}. A photograph of SBSL is shown in Fig. \ref{fig:sbsl}. In contrast to MBSL, there are no other bubbles present for the emitted light to scatter from. Similarly, the bubble does not interact with other bubbles and, due to its tiny extent, does not interact with the container walls. Both theory and experiments are greatly simplified compared to MBSL. Practically all progress on understanding SL has come from SBSL, with some authors even calling it ``the hydrogen atom of sonoluminescence" \cite{lohse2018bubble,crum1994sonoluminescence}. 

With this in mind, let's use SBSL as an idealization of shrimpoluminescence. The discussion will be much simpler if we further specialize to \emph{stable} SBSL. In stable SBSL the bubble is driven to produce light but does not disappear after collapsing; rather, it is periodically driven and emits light every cycle. This is in contrast to shrimpoluminescence, where the bubble is transient; i.e. it disappears after emitting light. The significance of this contradiction is addressed later.

In a SBSL\footnote{From here on,``SBSL" means \emph{stable} SBSL unless explicitly specified.} experiment, a bubble filled with noble gas is trapped at the center of a flask containing water [Fig. \ref{fig:summary_fig} (a)]. A transducer is connected to the flask and is tuned to excite one of the flask's vibrational modes. The flask's oscillations induce oscillating pressure in the fluid, $P(t)$, shown as a function of time in Fig. \ref{fig:summary_fig} (b). The oscillating pressure drives the bubble's expansion/contraction. For just the right driving amplitude and frequency, we get the complicated behavior of the bubble's radius, $R(t)$, shown in Fig. \ref{fig:summary_fig} (b). To understand how these dynamics lead to SBSL, let's look at the blown-up snapshots of the bubble at the times $t_1$, $t_2$ and $t_3$:
\begin{figure}
\centering
\includegraphics[width=1\linewidth]{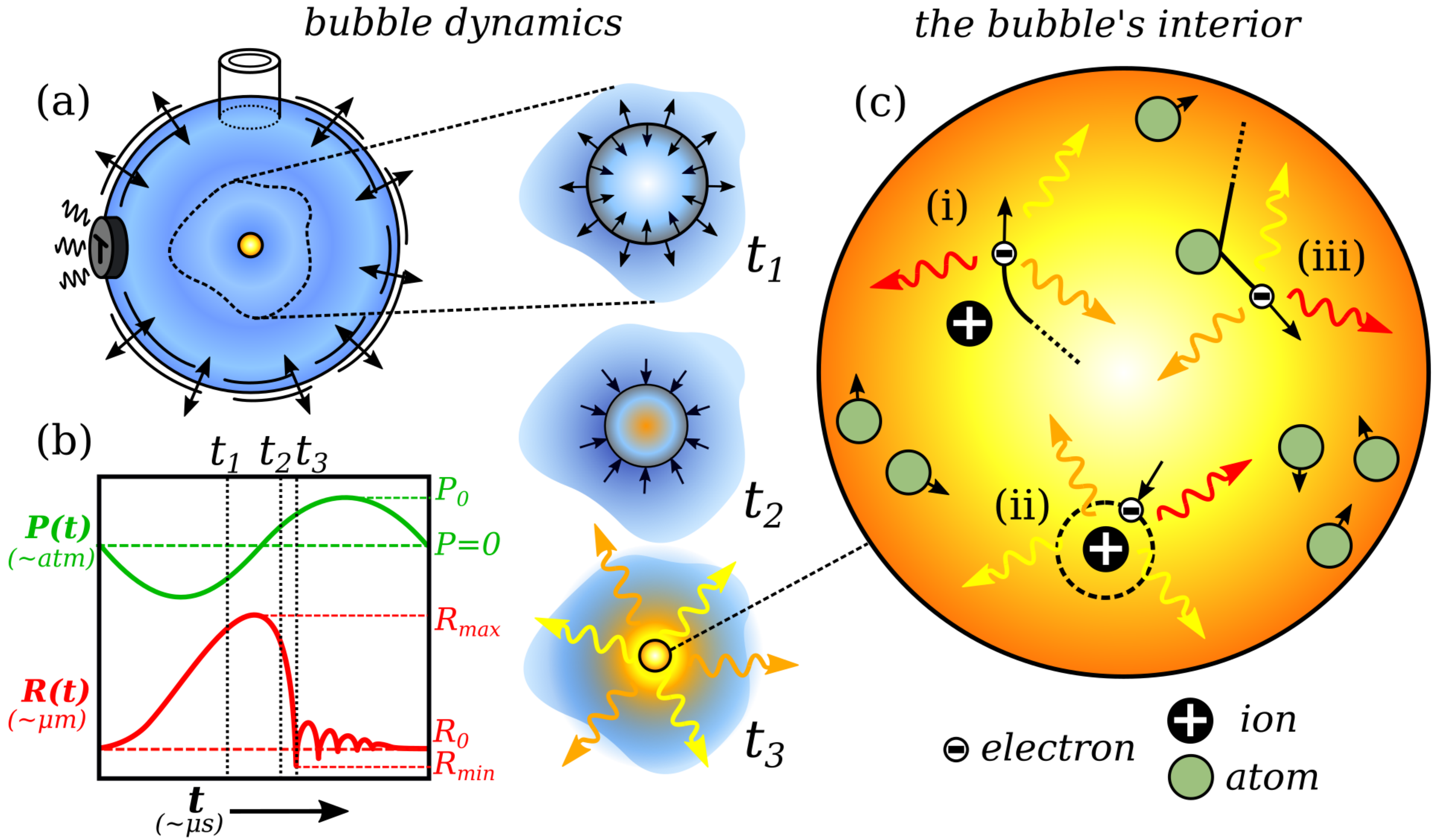}
    \caption{Schematic of the physics of SBSL. (a) Diagram of an SBSL experiment with a zoomed-in region around the bubble at different times shown to the right. A transducer, labeled as ``T", drives the flask and bubble. (b) Time-dependence of the driving pressure $P(t)$ and bubble radius $R(t)$  during SBSL. $P_0$ ($\sim 1.1-1.5$ atm) is the driving pressure amplitude measured from $P=0$. $R_0$ ($\sim 5\mu $m) is the ``ambient" radius of the bubble while $R_{min}$ ($\sim 0.5 \mu $m) and $R_{max}$ ($\sim 50 \mu $m) are, respectively, the bubble's minimum and maximum radii during SBSL. The time to compress the bubble from $R_{max}$ to $R_{min}$ is $\approx 1 \mu$s. (c) Inside the bubble during the light pulse at $t=t_3$. Electrons, ions, and atoms are labeled in the legend. The mechanisms of light emission are (i) electron-ion bremsstrahlung, (ii) electron-ion recombination, and (iii) electron-atom bremsstrahlung. }
\label{fig:summary_fig}
\end{figure}
\begin{itemize}
    \item [$\bm{t}$=$\bm{t_1}$] The driving pressure in the fluid is negative and tensile stresses cause the bubble to expand. The energy cost to break molecular bonds in the liquid opposes increasing the bubble's surface area. Surface-tension opposes the driving pressure and the bubble expands slowly.
    \item [$\bm{t}$=$\bm{t_2}$] The driving pressure is now positive and the bubble's radius is near its maximum. A large amount of work was done on the liquid to expand the bubble. The gas is dilute and offers little resistance to compression; nothing opposes accelerating the bubble's wall. Surface-tension cooperates with the positive driving pressure to convert the work back into kinetic energy. The bubble violently collapses in a process called \emph{cavitation}. Its radius shrinks from $R_{max}\sim 50~\mu$m to $R_{min} \sim 0.5~\mu$m, compressing the volume by a factor of $10^6$ in a few $\mu$s.
    \item [$\bm{t}$=$\bm{t_3}$] The bubble is at its minimum radius. During cavitation, the gas was compressed so quickly that heat could not flow out of the bubble. For a very short time ($\sim 0.003\% $ of the driving pressure's period \cite{suslick2008inside}), the temperature in the bubble reaches $10,000-30,000$ K \cite{an2006mechanism,hilgenfeldt1999simple,an2008spectral,an2009diagnosing}. The gas ionizes and emits light. After the light pulse, there is a period of ``after-bouncing" until the driving pressure becomes negative and the cycle begins again.
\end{itemize}

The interior of the bubble at $t=t_3$ is depicted in Fig. \ref{fig:summary_fig} (c). When the temperature is $\sim 10,000$ K, about $1 \%$ of the atoms in the gas ionize \cite{hilgenfeldt1999simple,yasui1999mechanism}, creating pairs of free electrons and positively charged ions. The Coulomb forces from ions deflect the electrons, changing their kinetic energy. The lost energy is converted into radiation through bremsstrahlung (``bremsstrahlung" means ``braking-radiation") \cite{zel2002physics,griffiths2005introduction,jackson1999classical}. \emph{Electron-ion bremsstrahlung} is labeled as (i) in Fig. \ref{fig:summary_fig} (c). Similarly, if an electron and an ion combine to form a neutral atom, light is emitted as their energies change: \emph{electron-ion recombination} is labeled as (ii). The degree of ionization is low, but the Coulomb force is long-ranged so electron-ion bremsstrahlung and electron-ion recombination are important \cite{zel2002physics}. One would ordinarily expect electron-atom scattering to be irrelevant since the electron-neutral particle interaction is very short-ranged. However, the pressure and particle density in the bubble at maximum compression are so large that electrons frequently collide with atoms, emitting radiation through \emph{electron-atom bremsstrahlung}. This mechanism is labeled as (iii) in Fig. \ref{fig:summary_fig} (c). Electron-ion bremsstrahlung, electron-atom bremsstrahlung, and electron-ion recombination are the most important mechanisms that produce light during SBSL \cite{hilgenfeldt1999simple,hilgenfeldt1999sonoluminescence,an2006mechanism,an2008spectral,an2009diagnosing,suslick2008inside}.

The goal of the rest of this paper is to introduce the theory needed to understand shrimpoluminescence in detail, though we will mainly focus on SBSL \footnote{It is apparently simpler to characterize SBSL without involving shrimp; convincing the shrimp to snap requires tickling them \cite{lohse2001snapping,versluis2000snapping,lohse2018bubble}. Besides notable exceptions \cite{tang2019bioinspired}, practically all work on SBSL has not involved shrimp.}. First, we will look at the problem of the fluid containing the bubble and derive an equation of motion for the bubble's wall. Next, we will use our results for the bubble's wall to understand the dynamics of the gas trapped in the bubble, calculating the temperature along the way. Finally, we will see how knowing the mechanisms of light emission/absorption in the bubble allow us to calculate the wave-length dependence of the bubble's light spectrum. Throughout the course of our journey, we will stick to the simplest results that still contain the essential physics we need. Connections to more advanced treatments, and their implications, are provided for completeness.

\section{Bubble Dynamics}
 
Lord Rayleigh solved the problem of a vapor filled cavity collapsing in water (the so-called Rayleigh collapse), giving the first rigorous theoretical treatment of cavitation in the early 1900's \cite{rayleigh1917pressure,plesset1977bubble}. He found that the bubble wall's velocity diverges, giving rise to ``cavitation." Since then, the theory of \emph{bubble dynamics} has been refined considerably and we will devote a large portion of this paper to it \cite{prosperetti1999old,plesset1977bubble,plesset1977bubble,brenner2002single,lofstedt1995sonoluminescing,barber1992resolving}. We split this into parts. First, we derive an equation of motion for the bubble's wall. Then we review some important experimental developments that allow us to use a simple model for the dynamics of the gas in the bubble. Finally, we compare results of simulated bubble dynamics to experimental data. The connection to experiment is what allows us to calculate the temperature in real bubbles.

\subsection{The Bubble Wall}
It turns out the dynamics in SBSL are quite well described both qualitatively and quantitatively by the classical theory of bubble dynamics \cite{prosperetti1999old,brenner2002single,prosperetti1986bubble,plesset1977bubble,suslick2008inside,yasui2018acoustic,brennen2014cavitation}. The most widely known modern work is that of Plesset \cite{plesset1977bubble,plesset1949dynamics,prosperetti1986bubble}, resulting in the so called ``Rayleigh-Plesset" (RP) equation which we are going to derive. The field of bubble dynamics is quite mature and is too large to review here; instead, the reader is referred elsewhere \cite{prosperetti1999old,brenner2002single,prosperetti1986bubble,brennen2014cavitation}. 

We start with some assumptions. The radius of a typical flask used in a SBSL experiment is $R_F\sim$ cm, while the radius of the bubble is $R\sim \mu$m. Then $(R/R_F)^3 \ll 1$ and we take $R_F\equiv R_\infty$, where the limit $R_F\rightarrow \infty$ is implied. Since the bubble is tiny compared to the fluid, we consider the dynamics in the fluid as if there is no bubble present. The bubble re-enters the theory as a boundary condition. We also assume spherical symmetry, so that the fluid's flow is purely radial. We will call the radial velocity of the fluid $u(r,t)$. The assumption that the bubble is always spherical seems like a rather drastic approximation but is validated experimentally \cite{prosperetti1999old,brenner2002single}. Finally, we assume the fluid is incompressible. Then the fluid density, $\rho$, is constant, and the flow into a region of space must exactly cancel the flow out. An important consequence is that flows across spherical shells at different radii must be equal: 
\begin{equation}
    \rho 4\pi R_1^2 \cdot u(R_1,t) = \rho 4\pi R_2^2 \cdot u(R_2,t). 
    \label{eq:flow}
\end{equation}

We will derive the RP equation from the principle of conservation of energy \cite{yasui2018acoustic,leighton2007derivation}. The mass of a spherical shell of fluid with thickness $dr$ and radius $r$ is $m=\rho 4 \pi r^2 dr$. The speed of the shell is $u(r,t)$ and it's kinetic energy is $dE_k(r)=1/2\cdot \rho 4\pi r^2 dr \cdot u^2(r,t)$. The total kinetic energy in the liquid outside the bubble is
\begin{equation}
    E_k=2\pi\rho\int_R^{R_\infty} r^2 u^2(r,t) dr = 2\pi\rho  \left( R^2 \dot{R} \right)^2 \int_R^{R_\infty} \frac{1}{r^2} dr = 2 \pi \rho R^3 \dot{R}^2
\label{eq:kinetic_energy}
\end{equation}
where we used the incompressibility eq. \ref{eq:flow}. The fluid's kinetic energy depends on the bubble's instantaneous radius, $R(t)$, and velocity, $\dot{R}$. The kinetic energy in the fluid has to equal the work done on the fluid by the bubble. We need to know the \emph{net-force} on the bubble wall to calculate the work. There are forces from surface-tension, fluid viscosity, pressure in the gas, and pressure in the liquid. Let's look at these in turn. 

If $\sigma$ is the energy per unit-area on the bubble's surface, then the total surface energy is $ 4\pi \sigma R^2$. The work required to change the bubble's radius by a small amount is $8 \pi \sigma R \cdot dR \equiv F_\sigma \cdot dR$. The force, $F_\sigma$, is uniform over the bubble's surface. The force per unit-area from surface-tension is $f_\sigma = 2\sigma /R$. The force per unit-area from viscous stress on a shell with radius $R$ is $f_\eta=2\eta \partial \dot{r}/ \partial r |_{r=R}$ \cite{yasui2018acoustic,prosperetti1999old,prosperetti1986bubble}. Using the incompressibility eq. \ref{eq:flow}, this gives $f_\eta=-4\eta \dot{R}/R$ on the bubble wall. 

Calculating the pressure in the gas, $p_g(t)$, is quite complicated and we will look at in detail later. For now, let's just assume $p_g(t)$ is a known quantity and move on to the pressure in the liquid. We can approximate the pressure in the liquid at the bubble to be equal to the pressure at the flask boundary, $p_\infty$. The pressure is $p_\infty=p_0+P(t)$ with $p_0$ the ambient, static pressure (usually $p_0=1$ atm) and $P(t)$ the time-dependent external pressure exerted on the liquid by the flask. In the case of SBSL, $P(t)$ is from the transducer driving the flask's oscillations. Usually the transducer is tuned to excite the lowest frequency resonance; it's wave-length, $\lambda \sim R_\infty$, is long so this is a good approximation \cite{prosperetti1986bubble,prosperetti1999old,plesset1977bubble}. For sinusoidal driving pressure, $P(t)=P_a \sin(\omega t)$. $P_a$ is the driving amplitude and $\omega$ is the driving frequency \cite{keller1980bubble}. Typically $P_a\sim1.1-1.5$ atm and $\omega\sim20-40$ kHz \cite{brenner2002single}. 

The net-force on the fluid at the bubble's wall is
\begin{equation}
    F_N=4\pi R^2 \left(p_g - \frac{1}{R}\left(2\sigma+4\eta \dot{R}\right) - p_\infty \right).
    \label{eq:net_force}
\end{equation}
The work done on the fluid by the bubble is $W=\int_{R_0}^R F_N dR'$. The lower bound, $R_0$, is a reference radius from which we measure the work done to expand the bubble to its instantaneous radius, $R=R(t)$. We arrive at an equation for conservation of energy in the fluid:
\begin{equation}
    2 \pi \rho R^3 \dot{R}^2= \int_{R_0}^{R} 4 \pi R'^2 \left(p_g -\frac{1}{R'}\left(2\sigma+4\eta \dot{R'}\right) - p_\infty \right) dR'
    \label{eq:E_conservation}
\end{equation}
This is almost what we need. To see what happens at the bubble's wall, we differentiate eq. \ref{eq:E_conservation} with respect to $R$. The derivative on the left side is tricky: $2 \pi \rho \partial_R \left(R^3 \dot{R}^2 \right)=6\pi \rho R^2 \dot{R}^2+4 \pi \rho R^3 \ddot{R}$. The derivative on the other side is easy. Finally, we arrive at the Rayleigh-Plesset equation:
\begin{equation}
    R\ddot{R}+\frac{3}{2}\dot{R}^2 = \frac{1}{\rho}\left[ p_g(t)-p_0-\frac{1}{R}\left(2\sigma+4\eta \dot{R} \right)-P_a\sin(\omega t) \right]
    \label{eq:RP}
\end{equation}
This is the equation of motion for the bubble's wall; solving it gives us the dynamics of the bubble as a function of time. The left hand side is the kinetic energy in the fluid due to motion of the bubble wall. The right hand side is potential energy; it has contributions from pressure in the gas, pressure in the liquid (driving pressure and static pressure), and from surface-tension and viscosity. The pressure in the gas always tends to expand the bubble, so it is positive. The sign of the driving pressure oscillates. Surface-tension always shrinks the bubble and viscosity always slows the bubble wall's speed; they are both negative. 

The RP equation eq. \ref{eq:RP} is pretty but its analytical solution is intractable. Direct solution of the RP equation is done numerically using Euler or Runge-Kutta methods \cite{yasui2018acoustic,yasui2015dynamics}. It is no great challenge computationally. For now, let's see if we can understand cavitation by making more approximations. Consider the interval shortly after the driving pressure becomes positive ($t_2\rightarrow t_3$ in Fig. \ref{fig:summary_fig}). The gas pressure is negligible and stresses from surface-tension and viscosity are all small. The bubble is collapsing, so we expect $\dot{R}^2$ to become very large \cite{rayleigh1917pressure,plesset1949dynamics,prosperetti1999old,brenner2002single}. If we neglect the right hand side of eq. \ref{eq:RP}, the RP equation reduces to Rayleigh's equation for a void collapsing in a fluid:
\begin{equation}
    -\ddot{R}=\frac{3}{2} \frac{\dot{R}^2}{R}.
    \label{eq:Rayleigh}
\end{equation}
The right hand side of eq. \ref{eq:Rayleigh} is strictly positive, so the acceleration is always negative. The bubble wall's \emph{speed} increases, causing larger negative acceleration, which continues to increase the speed... eventually the speed is infinite and there is cavitation. Eq. \ref{eq:Rayleigh} can be solved and the result is $R(t)\sim\left(1-t/\tau \right)^{2/5}$, with $\tau$ satisfying $R(\tau)=0$; the velocity, $\dot{R}\sim\left(1-t / \tau \right)^{-3/5}$, diverges. The same observations led to Rayleigh's explanation of cavitation in the early 1900's.

The \emph{physical} origin of cavitation in an incompressible fluid can be understood from eq. \ref{eq:flow} \cite{yasui2018acoustic}. The flow speed must scale as $u(R_1)=u(R_2)(R_2/R_1)^2$. Assume $R_2>R_1$ and that fluid is flowing from $R_2$ to $R_1$. More mass is contained in the shell with larger radius, so the speed of flow through the smaller shell must be larger by a factor of $\sim (R_2/R_1)^2$ for the density to be constant. As the flow approaches the origin, its speed diverges in order to ``make room" for in-coming fluid. 

In reality, the bubble's velocity is always finite and in the case of \emph{stable} SBSL, so is the radius. In the full RP equation (eq. \ref{eq:RP}) the most important term for slowing the bubble wall's motion is the diverging pressure inside the bubble \cite{brenner2002single}. If we had not assumed incompressibility, eq. \ref{eq:RP} would also include terms from the sound radiated by the bubble. A family of solutions, collectively called ``Rayleigh-Plesset equations," can be derived \cite{prosperetti1986bubble,prosperetti1988nonlinear,keller1956damping,lezzi1987bubble}. In these equations, it turns out that the most important term for slowing the bubble wall's diverging velocity is $\propto\dot{p}_g$. Adding only this term to the RP equation results in a popular variant that is only slightly more complicated than eq. \ref{eq:RP} \cite{lofstedt1995sonoluminescing,barber1997defining}. The observed error between solutions of this equation and experiment is only significant shortly after the Rayleigh collapse. It is in good quantitative agreement with measurements of the bubble's radius for the rest of the cycle \cite{brenner2002single}. 

Above, we assumed that the bubble remains spherical. Of course this isn't always true and \emph{shape-instabilities} occur for the right combinations of mechanical properties and driving pressure/frequency. Usually shape-instabilities destroy the bubble and kill SBSL so we won't focus on these issues here. We will always assume we are in a parameter-regime where SBSL occurs. The reader is referred to one of several reviews on the \emph{phase-space} of SBSL for details \cite{yasui2018acoustic,brenner2002single,hilgenfeldt1999sonoluminescence,an2009diagnosing}. 

\subsection{The Bubble's Interior}
Progress on understanding the bubble's \emph{interior} since the discovery of SBSL in the 1990's \cite{gaitan1992sonoluminescence} follows two paths \cite{brenner2002single,suslick2008inside,yasui2018acoustic}: (i) model calculations based on the RP equations and an equation of state for the pressure (from which we may calculate the temperature) are used to try to reproduce the easily measurable dynamics of the bubble. (ii) We model the light emission itself and compare our calculation to spectroscopic measurements of SBSL \cite{hiller1992spectrum,brenner2002single}. 
%Obviously these are related topics, but the distinction in the context of SBSL is important as we will now see.

Most early progress on the bubble's interior depended on the former method: if we predict the right dynamics, then we might know the correct pressure and temperature in the bubble. These data are then used to try to describe the light emission. Calculating the dynamics of the bubble's wall from one of the RP equations (e.g. eq. \ref{eq:RP}) requires the pressure inside the bubble as input. We need a suitable form for $p_g(t)$ in the RP equations that reproduces the measured radius-time curve $R(t)$ for a stable cavitating bubble. It turns out that this is a very complicated problem: gas diffusion between the bubble and liquid varies the number of particles present and, to make things worse, the conditions inside the bubble facilitate chemical reactions between the air and water-vapor, changing the properties of gas dynamically \cite{brenner2002single}. Brenner et. al elegantly summarized the significance of this problem \cite{brenner2002single}: ``one of the exciting features of modern research on SBSL is that it is a testing ground for how well mathematical models can deal with such a complicated situation."

\begin{figure}
\centering
\includegraphics[width=1\linewidth]{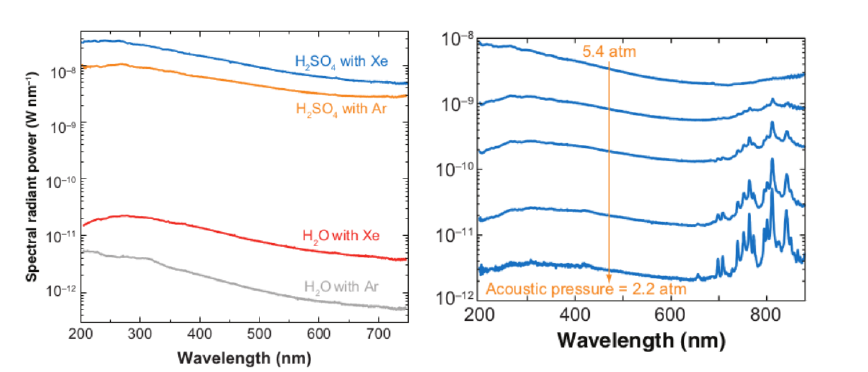}
    \caption{(left) Emission spectra as a function of wave-length, $\lambda$, for Ar and Xe bubbles in water and 85\% aqueous H$_2$SO$_4$. The pressure was optimized to give maximum intensity of emitted light. (right) Emission spectra as a function of wave-length, $\lambda$, for an Ar bubble in 85\% aqueous H$_2$SO$_4$ at different driving pressures, $P_0$. The spectral lines around $\sim 800$ nm are from Ar $4s\leftrightarrow4p$ transitions and disappear with increasing driving pressure. Adapted from ref. \cite{suslick2008inside}}
\label{fig:exp}
\end{figure}

The other way to understand the bubble's interior is its light spectrum. For the first decade or so, SBSL experiments were mainly done with \emph{air} bubbles in water \cite{suslick2008inside,brenner2002single,gaitan1992sonoluminescence}. Careful measurements of light emission from SBSL in water resulted in an almost featureless spectrum (see Fig. \ref{fig:exp}) and attempts to fit it as black-body radiation were not successful \cite{hiller1992spectrum}. Little progress was made on this front for quite some time \cite{brenner2002single}. The utility of analyzing the light spectrum of SBSL in water is succinctly described by Suslick \cite{suslick2008inside}: ``because of the inherent ambiguity associated with the analysis of featureless spectra of unknown origin, a more rigorous explanation [of SBSL] is unlikely to be generated". 

Some attempts were made to measure SBSL using non-aqueous liquids (e.g. alcohol, silicone oil) but experiments were not very successful \cite{weninger1995sonoluminescence,barber1997defining,barber1997defining}. Eventually, different gases were tried in water. Air is $\sim 80\%$ N$_2$, $\sim 20\%$ O$_2$, and $\sim 1\% $ Ar so, naturally, pure O$_2$ or N$_2$ were tried first. Degassed water \emph{regassed} with N$_2$, O$_2$, or even a mixture of O$_2$ and N$_2$ did not to produce SBSL \cite{hiller1994effect}. It was discovered that a noble gas was required for SBSL to occur \cite{barber1997defining,hiller1994effect,brenner2002single}. This resulted in the \emph{argon rectification hypothesis}, which claims that all species in the air inside the bubble besides Ar are gradually ejected until all that remains is pure Ar \cite{lohse1997sonoluminescing,brenner2002single,yasui2018acoustic,suslick2008inside}. The hypothesis is based on the fact that, at the high temperature inside the bubble, dissociation of O$_2$ and N$_2$ is possible. These species react with dissociated water vapor radicals to form new species that are soluble in water. As the pressure becomes very large during the compression stage of SBSL, the soluble materials leave the bubble and do not re-enter since their solubility in water is enormous compared to the Ar content of the bubble \cite{lohse1997sonoluminescing}. Over many cycles, the contents of an air bubble in water become nearly pure Ar. It was also realized that SL would be much more intense if the contents of the bubble are a pure inert gas: if the contents are e.g. molecules, the specific heat of the gas is larger, resulting in lower temperature and decreased light production from the bubble \cite{brenner2002single,yasui2018acoustic,lohse2018bubble,suslick2008inside}. This mechanism is used to explain the relatively weak light observed in MBSL: the bubbles are transient and cannot eject a significant amount of O$_2$ or N$_2$ over a single cycle. The current belief is that the contents of a bubble undergoing SBSL in water are (nearly) pure Ar after $\sim 10^3$ cycles \cite{brenner2002single,yasui2018acoustic}.  

Most studies continued to use water as the host liquid until it was realized that SBSL in aqueous H$_2$SO$_4$ produces light $10^3$ times brighter than in water, allowing more precise measurements of the light spectrum \cite{flannigan2005plasma,flannigan2005plasma1,flannigan2006measurement}. The mechanical properties in aqueous H$_2$SO$_4$ allow stable SBSL with larger bubble radii than in water, increasing the emitting volume. More importantly, new measurements revealed the phase space of SBSL in aqueous H$_2$SO$_4$ included a much larger range of driving pressure than water [Fig. \ref{fig:exp} (right)]. SBSL in aqueous H$_2$SO$_4$ could be driven at very low pressure, leading to much lower temperature bubbles, while driving at large pressure led to conditions similar to SBSL in water. 

Careful experiments revealed that the spectrum depended critically on the noble gas content and driving pressure. Suslick et. al measured the emission spectra from Ar, Xe, and Kr bubbles in aqueous H$_2$SO$_4$ \cite{flannigan2005plasma,flannigan2005plasma1,flannigan2006measurement,suslick2008inside}. At low driving pressure, they identified spectral lines from Ar$^+$, Xe$^+$, and Kr$^+$ excited state transitions, proving that the core of the bubble is plasma. Perhaps equally as crucially, they found the absence of emission lines from components of aqueous H$_2$SO$_4$ vapor at large driving pressure \cite{suslick2008inside,flannigan2006measurement,flannigan2006measurement}: the plasma contained only the noble gas, experimentally confirming the argon rectification hypothesis. This work also explained why the spectrum in water is featureless; the large pressure in SBSL in water results in an increased collision rate between particles, broadening the spectral lines until they are no longer visible  \cite{an2009diagnosing,suslick2008inside,flannigan2005plasma,flannigan2005plasma1,flannigan2006measurement}.

The argon rectification hypothesis greatly helped to refine models for the gas dynamics in the bubbles and confirmed why simple models usually work well \cite{suslick2008inside,brenner2002single}. With this in mind, we will only focus on a simple model for the gas dynamics in this paper. Modern reviews on modelling the bubble's interior in SBSL exist elsewhere \cite{brenner2002single,yasui2018acoustic,brennen2014cavitation}. 

\subsection{Gas Dynamics}
The most straightforward way to model the gas dynamics in the bubble is through direct solution of the Navier-Stokes equations for the gas \cite{brenner2002single}. In fact, the most accurate quantitative predictions of the conditions in the bubble's interior take this path \cite{an2006mechanism,an2008spectral,an2009diagnosing,flannigan2005plasma,flannigan2005plasma1,flannigan2006measurement}. Equations of motion with varying degrees of sophistication are derived and the coupled system of gas dynamical and RP equations are solved numerically. As already mentioned, this is a very complicated procedure if we want to account for mass and heat transfer across the bubble wall \cite{brenner2002single,yasui2018acoustic}. Instead, we will use our knowledge of the bubble's contents to argue for a simple model that qualitatively describes the right dynamics. We will discuss how it can be extended to more accurately represent the dynamics later.

Recalling that the bubble is (almost) purely noble gas, a simple but reasonable model for the gas dynamics, at least during the collapse, is that of an adiabatically, quasistatically compressed Van der Waals gas \cite{brenner2002single,lofstedt1995sonoluminescing,barber1997defining,lofstedt1993toward,hilgenfeldt1999simple}. In the context of SBSL, the interaction term in the Van der Waals (VdW) equation of state is usually neglected and the only modification to the ideal gas result is the excluded volume of the real gas molecules, $V_h = 4 \pi h^3/3$, with $h$ the VdW \emph{hard-core radius}. The pressure in an adiabatically compressed \emph{ideal} gas at volume $V_2$ is \cite{schroeder1999introduction}
\begin{equation}
    P_2 = P_1 \left(\frac{V_1}{V_2}\right)^\gamma
\label{eq:P}
\end{equation}
where $P_1$ is the initial pressure at volume $V_1$ and $\gamma=C_P/C_V$ is the ratio of the constant-pressure and constant-volume specific heats respectively. In our case, we assume that our bubble has an ambient radius, $R_0$, where the pressure is at its equilibrium value, $P_0$. Usually $R_0$ has to be determined from an experiment, but let's assume we know it for now. The ambient volume of the bubble is $V_0 =  4 \pi R_0^3/3$. It follows from our discussion of the RP equation that $P_0=p_0+2\sigma /R_0$. We want to calculate the pressure in the gas, $p_g(t)$, as a function of the bubble's radius, $R(t)$, determined from the RP equation. Plugging these data into eq. \ref{eq:P}, and subtracting the excluded volume, we arrive at a very commonly used equation in the context of SBSL \cite{brenner2002single,lofstedt1995sonoluminescing,barber1997defining,lofstedt1993toward,hilgenfeldt1999simple,sivasubramanian2002temperature}:
\begin{equation}
    p_g(t) = \left( p_0+2\frac{\sigma}{R_0} \right) \left[ \frac{R_0^3-h^3}{R^3(t)-h^3} \right] ^ \gamma
    \label{eq:p_g}
\end{equation}
It's clear that the purpose of the excluded volume is to make sure that if the bubble collapses so strongly that its contents become incompressible (i.e. $R(t)\rightarrow h$), the pressure diverges \cite{lofstedt1993toward,brenner2002single}. For pure Ar, $h\approx R_0/8.86$ \cite{hilgenfeldt1999sonoluminescence}. Typically $R_0\sim \mu$m and the minimum radius after cavitation is $R_{min} \approx 0.2 \cdot R_0$. For $P_0 \approx10^5$ Pa (1 atm), the peak pressure in the gas is $\sim 10^{9}$ Pa.

Physical intuition tells us why eq. \ref{eq:p_g} is sensible during the bubble's collapse: the bubble wall's velocity is fast and very little heat can flow out during compression. During the expansion and after-bounce stages of the bubble's cycle, which comprise its majority, the gas dynamics should instead be regarded as \emph{isothermal} \cite{brenner2002single}. This is true because, when the wall's velocity is comparable to the heat diffusion timescale, the temperature throughout the bubble is nearly equal to that in the liquid \cite{prosperetti1999old,brenner2002single,yasui2018acoustic}. The relevant modification to eq. \ref{eq:p_g} is replacing $\gamma\rightarrow 1$. A neat way of including both isothermal and adiabatic gas dynamics is continuously varying the exponent, $\gamma$ in eq. \ref{eq:p_g}, between its adiabatic and isothermal values, $C_P/C_V$ and $1$, respectively. This technique has been commonly used in calculations of SBSL in water \cite{hilgenfeldt1999simple,hilgenfeldt1999sonoluminescence,prosperetti1986bubble,brenner2002single}. Simple forms, e.g. eq. \ref{eq:p_g} with the constant adiabatic exponent $\gamma=5/3$, give close to the correct bubble dynamics with the most obvious error occurring during the isothermal after-bounces (see Fig. \ref{fig:mie}). 

%more advanced solutions of the gas Navier-Stokes equations improving only on estimates for the conditions inside the bubble \cite{brenner2002single,yasui2018acoustic}.

Recalling that one of main goals of modeling the bubble dynamics in SBSL was to calculate the temperature inside the bubble, we can also derive an equation for $T(t)$ \cite{barber1997defining,brenner2002single,sivasubramanian2002temperature}:
\begin{equation}
    T(t) = T_0 \left( \frac{R_0^3-h^3}{R^3(t)-h^3} \right)^ {\gamma-1}
    \label{eq:T(t)}
\end{equation}
We said that for most of the bubble's cycle, the gas dynamics is isothermal: $T_0$ is the temperature of the liquid. For $T_0\approx 300$ K, and using the radii above, the peak temperature is $T_{max}\sim 10,000$ K. The only ``unknown" quantity still preventing us from actually calculating the bubble's dynamics and temperature is $R_0$ in eqs. \ref{eq:p_g} and \ref{eq:T(t)}. We learn how to determine $R_0$ in the next section.

\subsection{Bubbles in the Lab}

%\begin{figure}
%\includegraphics[width=0.9\linewidth]{figs/flask.pdf}
%    \caption{(a) Schematic of a spherical ``acoustic levitation cell". The piezoelectric transducers that drive the bubble trapped in the flask are labeled in the diagram. (b) Photograph of a sonoluminescing bubble in 85\% aqueous H$_2$SO$_4$. (c) Schematic of a Mie scattering experiment. Panels (a), (b), and (c) are from refs. \cite{brenner2002single}, \cite{suslick2008inside}, and \cite{gompf2000mie} respectively.}
%\label{fig:flask}
%\end{figure}

\begin{figure}
\centering
\includegraphics[width=0.8\linewidth]{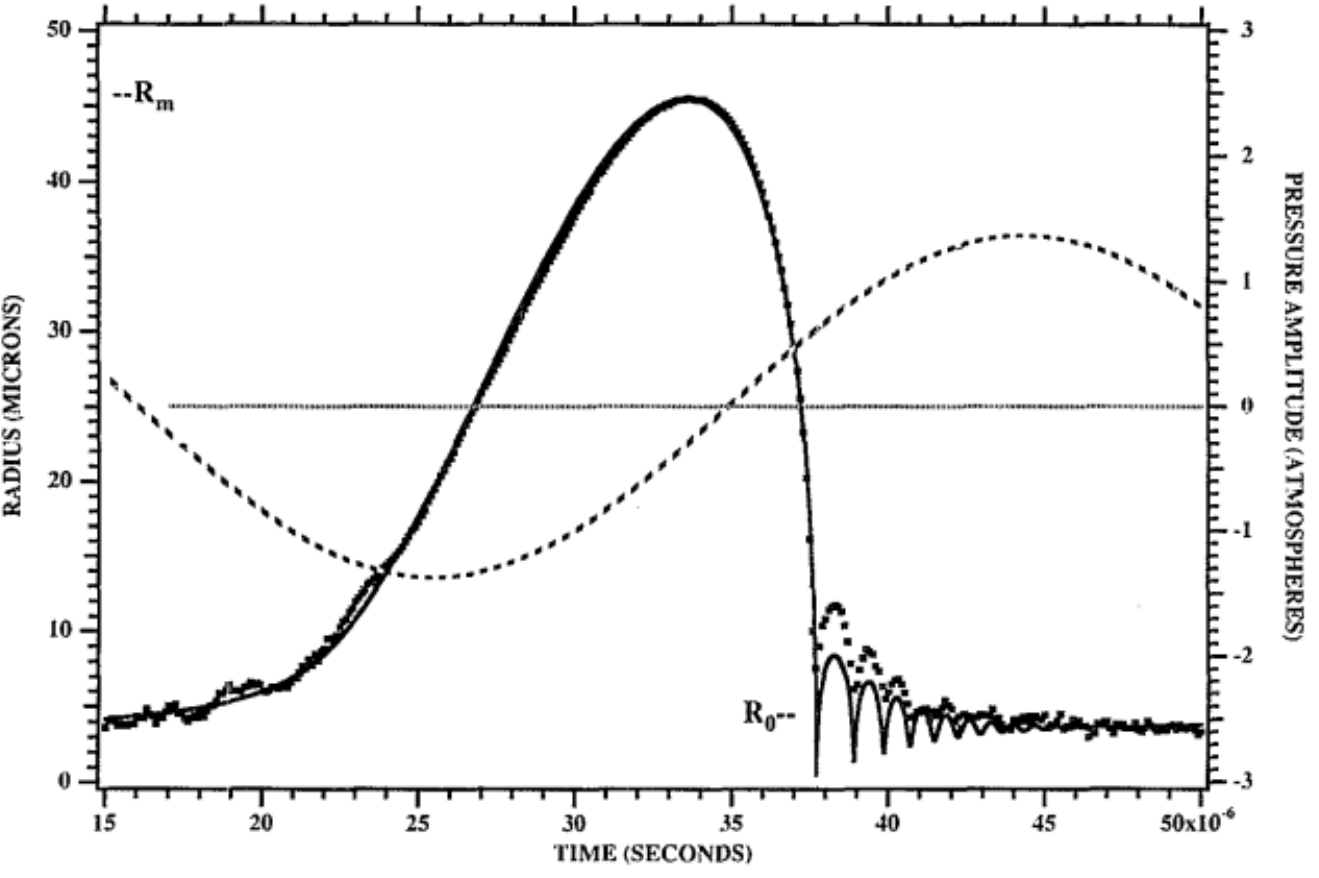}
    \caption{Numerical solution of the $\propto \dot{p}_g$ RP equation variant discussed earlier with the gas dynamics given by the simple form in eq. \ref{eq:p_g}. The solid line is the result of the calculation. The dots are experimental results from Mie scattering. The fit resulted in $R_0=4.5$ $\mu m$ and max temperature $8,500$ K. Adapted from ref. \cite{lofstedt1993toward}.}
\label{fig:mie}
\end{figure}

Creating stable, single bubbles in an experiment is not particularly challenging and can be done with standard and low-cost materials. A typical experimental setup is shown in Fig. \ref{fig:summary_fig} (a). Since we are concerned with explaining the light emission process itself, we will only sketch the experimental setup; it can be summarized as follows \cite{lentz1995mie,gaitan1990experimental,gaitan1992sonoluminescence,gompf2000mie,brenner2002single,yasui2018acoustic,brennen2014cavitation,suslick2008inside}. A suitable sample of liquid is placed into a flask. Stuck to the outside are piezoelectric transducers driven sinusoidally to excite a resonance of the flask: a typical flask is a few cm across with a resonance at $\approx 20$ kHz \cite{brenner2002single}. The frequency is chosen to form a pressure \emph{antinode} at the center of the flask, trapping a bubble there.

The driving pressures relevant to SBSL ($\sim 1.1-1.5$ atm) are too small to cause bubbles to form spontaneously \cite{brenner2002single}. Instead, a bubble is usually \emph{seeded} somehow. In the original work of Gaitan, an air bubble was injected using a syringe \cite{gaitan1992sonoluminescence}. More recently, seeding methods involve shooting a small jet of water into the flask (like our shrimp friends!) or blasting the water with a laser to boil a small volume \cite{suslick2008inside,yasui2018acoustic}. The later method is preferred as it enables more precise experimental control over the bubble's ambient radius, $R_0$.

Now let's discuss two experimentally accessible quantities that have turned out the be important. (i) The bubble's radius may be measured by \emph{Mie scattering} and compared directly to solutions of the RP equations, allowing us to ``measure" $T(t)$. (ii) The light can be analyzed with a spectrometer, giving us information on the bubble's contents: in principle, we can learn about temperature and pressure inside the bubble \cite{hiller1992spectrum,hilgenfeldt1999simple,flannigan2005plasma,flannigan2005plasma1,flannigan2006measurement}. We look at this in the next section.

In Mie scattering, a laser is shot at a bubble and the scattered intensity is analyzed to determine $R(t)$ \cite{matula1999inertial,lentz1995mie}. Modeling the bubble as a homogeneous dielectric sphere \cite{jackson1999classical}, the angular distribution of light scattered from the bubble at \emph{fixed radius} and the time-dependent scattering off the bubble at \emph{fixed angle} allow use to accurately measure the bubble's radius as a function of time \cite{brenner2002single}. Results from a Mie scattering experiment are shown as dots in Fig. \ref{fig:mie}. Parameters used to calculate the gas dynamics in the bubble are fit to the measured radius to complete the model of the bubble dynamics. For the gas model in the last section, eq. \ref{eq:p_g}, the parameter to fit is $R_0$. Once we fit it, we know the temperature in the bubble from eq. \ref{eq:T(t)}. A calculated curve is shown in Fig. \ref{fig:mie}. The dynamics fit the experiment well. $R_0=4.5~ \mu$m and the ``measured" temperature is $8,500$ K. 

Before discussing the light spectrum, we note that it is possible to calculate the temperature from ``first-principles" without fitting to experiment. This can done by direct numerical solution of the gas's Navier-Stokes equations. The results are qualitatively similar to simple model calculations, but are in slightly better quantitative agreement when used to calculate the light spectrum \cite{an2009diagnosing,an2008spectral,an2006mechanism}. 

\section{Let there be light!}

Now that we know the temperature in the bubble, $T(t)$, we are finally ready to discuss the light emission. A requirement for models of light emission in SBSL is \emph{local thermodynamic equilibrium} (LTE). LTE is expected to be true in a bubble undergoing SBSL since the particle density ($\sim 10^{28}/m^3$) and temperature ($\sim 10^4$ K) are very large, causing collisions between particles to occur so frequently that equilibrium is guaranteed \cite{brenner2002single,hilgenfeldt1999sonoluminescence,yasui1999mechanism}. 

Any ordinary matter at finite temperature will continuously and spontaneously emit radiation. In LTE, the rate energy is emitted equals the rate it is absorbed \cite{zel2002physics}. This means that we only need to know \emph{either} the emission coefficient \emph{or} the absorption coefficient of a material to know the intensity that it emits radiation. The intensity, $I(\lambda,T)$, is the amount of energy passing per unit-time and unit-solid angle through a unit-area oriented perpendicular to the light's propagation direction. An ideal emitter/absorber is called a ``black-body."  For a black-body, the intensity is given by Planck's law: \cite{schroeder1999introduction}
\begin{equation}
    I_{B}(\lambda,T(t))=\frac{2 h c^2}{\lambda^5}\left[\exp\left(\frac{hc}{\lambda k_B T(t)}\right)-1\right]^{-1}
\end{equation}
The subscript $B$ is to remind us this is for a black-body. $\lambda$ is the wave-length of the light, $c$ is the speed of light in vacuum (we assume the index of refraction is $n\approx 1$ \cite{hilgenfeldt1999simple,hilgenfeldt1999sonoluminescence,zel2002physics}), $h$ is Planck's constant, and $k_B$ is Boltzmann's constant. We can use the temperature, $T(t)$, from eq. \ref{eq:T(t)}, to calculate the instantaneous intensity radiated in the bubble. Integrating over the bubble's volume and over time, we can calculate the number of photons produced during SBSL; a quantity that can be directly compared to experiment \cite{hilgenfeldt1999simple,hilgenfeldt1999sonoluminescence,an2006mechanism,an2008spectral,an2009diagnosing}. Calculations that assume the bubble is a black-body over-estimate the number of photons by orders of magnitude \cite{hilgenfeldt1999simple,brenner2002single}. We need to use a model that includes finite opacity.

Let's call the wave-length and temperature dependent absorption coefficient $\kappa(\lambda,T)$. For a ray of light emitted somewhere in the bubble, the intensity after travelling a distance $s$ is \cite{zel2002physics,hilgenfeldt1999simple,taylor1969experimental}
\begin{equation}
    I(\lambda,T(t))=I_{B}(\lambda,T(t))  \left[ 1-\exp \left( -\kappa[\lambda,T(t)] s \right) \right].
    \label{eq:intensity}
\end{equation}
We are assuming that the temperature in the bubble is spatially uniform so that $\kappa$ is too \cite{hilgenfeldt1999sonoluminescence,hilgenfeldt1999simple}. In the limit that $\kappa \rightarrow\infty$, i.e. an opaque bubble, we recover the result for a black-body emitter. On the other hand, for a nearly transparent bubble, $\kappa \ll 1$ and $I\approx I_B \kappa s$. The number of photons emitted is greatly reduced; a result that what we require. A great deal of work has been devoted to identifying the relevant absorption mechanisms leading to $\kappa(\lambda,T)$ \cite{hilgenfeldt1999simple,brenner2002single,hilgenfeldt1999sonoluminescence,yasui1999mechanism,flannigan2005plasma,flannigan2005plasma1,flannigan2006measurement,suslick2008inside,an2009diagnosing,an2008spectral,an2006mechanism}. 

\begin{figure}
\centering
\includegraphics[width=1\linewidth]{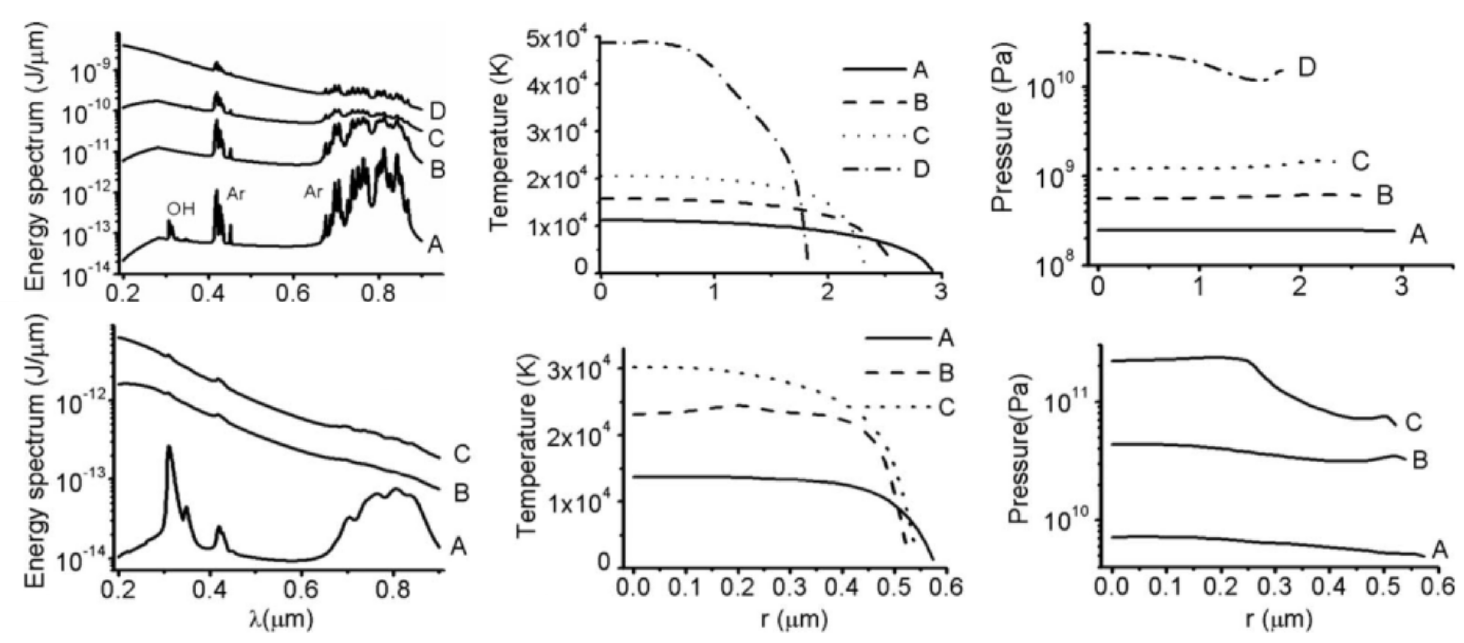}
    \caption{Simulations of an Ar bubble in 85\% aqueous H$_2$SO$_4$ (top-row) and in water (bottom-row). The gas Navier-Stokes equations were solved directly. The $1^{st}$ column shows the emitted spectrum as a function of wave-length, $\lambda$. The $2^{nd}$ and $3^{rd}$ columns shows the temperature and pressure as a function of position inside the bubble at the time the bubble is compressed to its minimum size. In the top row, the liquid temperature was 20 $^o$C and the bubble was driven at $P_0$=1.5, 1.7, 2.0, and 4.0 atm for curves A, B, C, and D respectively. In the bottom row, A and B are driven at $P_0=$1.22 and 1.32 atm, respectively, at 20 $^o$C. Curve C is at $P_0$=1.32 atm and 0 $^0$C. All data are from adapted from ref. \cite{an2009diagnosing}.}
\label{fig:light}
\end{figure}

Let us first discuss SBSL of a \emph{pure} Ar bubble in water. Due to the absence of lines in SBSL spectra, early models only considered \emph{continuous} emission processes \cite{hilgenfeldt1999simple,hilgenfeldt1999sonoluminescence}. The derivations of the formulae for absorption coefficients are too long to fit in this paper, so we will simply highlight what is important; we include one of them as an example. At the very high temperatures that occur in the bubble ($\sim10,000-30,000$ K, see Fig. \ref{fig:light}), it is possible for electrons to become ionized from their atoms \cite{zel2002physics}. When a free electron passes near an ion, it is deflected by the Coulomb force. The electron is accelerated, changing its kinetic energy. Since the electron is charged, it emits radiation through bremsstrahlung \cite{jackson1999classical,griffiths2005introduction,zel2002physics}. A \emph{classical} expression for absorption due to electron-ion bremsstrahlung is derived in ref. \cite{zel2002physics} and is frequently used in the context of SBSL \cite{hilgenfeldt1999simple,hilgenfeldt1999sonoluminescence,an2006mechanism,an2008spectral,an2009diagnosing}:
\begin{equation}
    \kappa(\lambda,T(t)) = \frac{4}{3} \left( \frac{2\pi}{3 m_e k_B T(t)} \right)^{1/2} \frac{Z^2 e^6 \lambda^3}{ (4 \pi \epsilon_0)^3 h c^4 m_e } N_+ N_e .
    \label{eq:absorption_coeff}
\end{equation}
In eq. \ref{eq:absorption_coeff}, $m_e$ is the electron rest mass, $e$ is the elementary charge, $Z$ is the ``effective" charge of the ion (usually taken to be $Z=1$), $\epsilon_0$ is the vacuum permittivity, and $N_+$ and $N_e$ are, respectively, the number density of ions and electrons. Usually $N_+=N_e$ and $N_+N_e=N_e^2$. The fact that $\kappa \propto N_e N_+$ is intuitive: the amount of energy emitted/absorbed by electrons scattering off ions ought to be proportional to the number of electron and ions present. $N_e$ is easily calculated from the ``Saha equation" of astrophysics \cite{hilgenfeldt1999simple,an2006mechanism,zel2002physics}. Quantum mechanical calculations of $\kappa$ for electron-ion bremsstrahlung have also been done (see ref. for a review \cite{eckert2015aether}), but the formulae are considerably more complicated and are seldom used in the context of SBSL. If the electron becomes bound instead of being deflected, a different expression for the absorption has to be used \cite{zel2002physics}. The coefficient depends on which Ar atomic level the electron is in, but the free electron energy is continuous and electron-ion \emph{recombination} radiation is continuous, consistent with the spectrum measured in water \cite{hilgenfeldt1999simple,hilgenfeldt1999sonoluminescence}. 

We're not done. It turns out that at temperatures relevant to SBSL ($\sim 10^4$ K), the fraction of atoms that are ionized is about $1\%$. The pressures that occur in SBSL ($\sim10^{12}$, see Fig. \ref{fig:light}) are so large that collisions between electrons and neutral atoms (which are otherwise rare) are frequent enough to be as important as electron-ion collisions \cite{zel2002physics,hilgenfeldt1999simple,frommhold1998electron}. A classical formula for the absorption coefficient for electron-atom bremsstrahlung is available in ref. \cite{zel2002physics}, though more recent works \cite{an2006mechanism,an2008spectral,an2009diagnosing} use numerical results from quantum mechanical calculations \cite{geltman1973free}. 

Eq. \ref{eq:absorption_coeff}, combined with analogous formulae for electron-atom bremsstrahlung \cite{geltman1973free} and electron-ion recombination \cite{zel2002physics} were used to accurately calculate the spectrum of SBSL from Ar in water \cite{hilgenfeldt1999simple,hilgenfeldt1999sonoluminescence,hammer2002light,yasui2001effect,yasui1999mechanism}. The only remaining error was small disagreement in the number of photons emitted \cite{an2006mechanism}, which was later accounted for after SBSL was discovered in aqueous H$_2$SO$_4$ \cite{flannigan2005plasma,flannigan2005plasma1,flannigan2006measurement,suslick2008inside}. This later work discovered that emission \emph{lines} for Ar$^+$ and Ar excited state transitions [$\sim 400$ nm and $\sim 800$ nm, respectively, in Fig. \ref{fig:light} (left)] and continuous emission from electron-OH$^{+}$ recombination are present. However, at the conditions encountered in SBSL in water [Fig. \ref{fig:exp} (right)], the emission lines are concealed by the dominant continuum emission processes at high temperature while the electron-OH$^{+}$ recombination contribution diminishes as the remaining vapor content is driven out of the bubble at high pressure \cite{suslick2008inside,flannigan2005plasma,flannigan2005plasma1,flannigan2006measurement}. More recent calculations of SBSL in both water and aqueous H$_2$SO$_4$ that include all of these mechanisms are in excellent agreement with experiment, as can be seen by comparing Fig. \ref{fig:exp} and Fig. \ref{fig:light}. 

\section{Summary and Outlook}
We now have everything we need to describe SBSL will-nilly! We can use the RP equation, eq. \ref{eq:RP}, or one of it's variants to calculate the radius time curve, $R(t)$, of a bubble in SBSL. With a suitable model for the gas dynamics in the bubble (e.g. eq. \ref{eq:p_g}) the coupled RP and gas equations are solved numerically and fit to measured $R(t)$. This procedure allows us to ``measure" the temperature in the bubble. With $T(t)$, and knowing the correct mechanisms for emission/absorption in the bubble (which are electron-ion bremsstrahlung, electron-atom bremsstrahlung, and electron-ion recombination), we can calculate the radiated intensity as a function of time. The results of this procedure are the wave-length dependence of the emitted radiation, the pulse width (by solving as a function of time), and the number of photons (by integrating over the bubble and over time). The fact that all of these quantities are in excellent agreement with experiments has led to the popular opinion that SBSL is a ``solved" problem \cite{yasui2018acoustic,lohse2018bubble}.

It is interesting to note other distinct models for SBSL. A class of ``fractoluminescence" models have persisted in spite of the success of the work presented in this paper. These are based on the hypothesis that shape instabilities cause the bubble to violently disintegrate, resulting in velocities faster than the fluid can flow \cite{prosperetti1997new}. The fluid ``fractures" and plasma forms during dielectric breakdown across the fracture, emitting light \cite{borisenok2020mechanisms,borissenok2008sonoluminescence}. These models haven't gained much traction, however, since the requirement that the bubble disintegrates contradicts the fact that they don't...

An even more exotic theory supposes SBSL is from ``quantum vacuum radiation" \cite{schwinger1992casimir1,schwinger1993casimir1}. Schwinger proposed that shrinking a cavity would change the number of electromagnetic modes confined in it, changing the energy expectation value; the lost energy would be carried away by photons. Direct calculation of the number of photons was carried out by others \cite{eberlein1996sonoluminescence}. It was found that the number of emitted photons would be $10^{-10}$ times too small \cite{unnikrishnan1996comment,lambrecht1997comment,garcia1997comment,milton1996comment,liberati2000sonoluminescence}. The quantum vacuum radiation theory has been abandoned.

The theory discussed in this paper most likely explains \emph{shrimpoluminescence}, though some discrepancies should be addressed. The foremost being that shrimpoluminescence is too dim to be seen by the naked eye \cite{lohse2001snapping},  while SBSL in the lab is not. Recall that shrimpoluminescence is transient, while the theory developed in this paper is for stable bubbles. For a transient bubble, it is not possible for air molecules or water vapor to be ejected from the bubble, a process which occurs over $\sim10^3$ cycles. These extra degrees of freedom prevent the temperature from becoming as large as in \emph{stable} SBSL and the amount of emitted light is suppressed.
Another issue is that the bubble produced by a snapping shrimp is not stationary \cite{lohse1997sonoluminescing,versluis2000snapping}. Translating bubbles are not quite spherical \cite{brenner2002single}, and small shape distortions could decrease the peak pressure and temperature in the bubble, suppressing the amount of light produced \cite{matula2000single}. Still, the essential physics we have learned from SBSL in this paper gives us the tools we need to \emph{shine light} on shrimpoluminescence. 

%\raggedbottom
\printbibliography

@article{liberati2000sonoluminescence,
  title={Sonoluminescence as a QED vacuum effect: probing Schwinger's proposal},
  author={Liberati, Stefano and Visser, Matt and Belgiorno, F and Sciama, DW},
  journal={Journal of Physics A: Mathematical and General},
  volume={33},
  number={11},
  pages={2251},
  year={2000},
  publisher={IOP Publishing}
}

@article{schwinger1993casimir1,
  title={Casimir light: a glimpse.},
  author={Schwinger, Julian},
  journal={Proceedings of the National Academy of Sciences},
  volume={90},
  number={3},
  pages={958--959},
  year={1993},
  publisher={National Acad Sciences}
}

@article{schwinger1992casimir1,
  title={Casimir energy for dielectrics.},
  author={Schwinger, Julian},
  journal={Proceedings of the National Academy of Sciences of the United States of America},
  volume={89},
  number={9},
  pages={4091},
  year={1992},
  publisher={National Academy of Sciences}
}

@article{eberlein1996sonoluminescence,
  title={Sonoluminescence as quantum vacuum radiation},
  author={Eberlein, Claudia},
  journal={Physical Review Letters},
  volume={76},
  number={20},
  pages={3842},
  year={1996},
  publisher={APS}
}

@article{lohse2001snapping,
  title={Snapping shrimp make flashing bubbles},
  author={Lohse, Detlef and Schmitz, Barbara and Versluis, Michel},
  journal={Nature},
  volume={413},
  number={6855},
  pages={477--478},
  year={2001},
  publisher={Nature Publishing Group}
}

@article{versluis2000snapping,
  title={How snapping shrimp snap: through cavitating bubbles},
  author={Versluis, Michel and Schmitz, Barbara and Von der Heydt, Anna and Lohse, Detlef},
  journal={Science},
  volume={289},
  number={5487},
  pages={2114--2117},
  year={2000},
  publisher={American Association for the Advancement of Science}
}

@article{brenner2002single,
  title={Single-bubble sonoluminescence},
  author={Brenner, Michael P and Hilgenfeldt, Sascha and Lohse, Detlef},
  journal={Reviews of modern physics},
  volume={74},
  number={2},
  pages={425},
  year={2002},
  publisher={APS}
}

@article{flannigan2005plasma1,
  title={Plasma formation and temperature measurement during single-bubble cavitation},
  author={Flannigan, David J and Suslick, Kenneth S},
  journal={Nature},
  volume={434},
  number={7029},
  pages={52--55},
  year={2005},
  publisher={Nature Publishing Group}
}

@article{lohse2018bubble,
  title={Bubble puzzles: From fundamentals to applications},
  author={Lohse, Detlef},
  journal={Physical review fluids},
  volume={3},
  number={11},
  pages={110504},
  year={2018},
  publisher={APS}
}

@article{borisenok2020mechanisms,
  title={On the Mechanisms of Sonoluminescence in Polar and Nonpolar Liquids},
  author={Borisenok, VA and Sedov, S Yu},
  journal={Physics of Atomic Nuclei},
  volume={83},
  number={11},
  pages={1575--1584},
  year={2020},
  publisher={Springer}
}

@article{prosperetti1997new,
  title={A new mechanism for sonoluminescence},
  author={Prosperetti, Andrea},
  journal={The Journal of the Acoustical Society of America},
  volume={101},
  number={4},
  pages={2003--2007},
  year={1997},
 publisher={Acoustical Society of America}
}

@article{borissenok2008sonoluminescence,
  title={Sonoluminescence: Two sources of light},
  author={Borissenok, VA},
  journal={Physics Letters A},
  volume={372},
  number={19},
  pages={3496--3500},
  year={2008},
  publisher={Elsevier}
}

@phdthesis{gaitan1990experimental,
  title={An experimental investigation of acoustic cavitation in gaseous liquids},
  author={Gaitan, Dario Felipe},
  year={1990},
  school={The University of Mississippi}
}

@article{crum1994sonoluminescence,
  title={Sonoluminescence},
  author={Crum, Lawrence A and Roy, Ronald A},
  journal={Science},
  volume={266},
  number={5183},
  pages={233--234},
  year={1994},
  publisher={American Association for the Advancement of Science}
}

@article{gaitan1992sonoluminescence,
  title={Sonoluminescence and bubble dynamics for a single, stable, cavitation bubble},
  author={Gaitan, D Felipe and Crum, Lawrence A and Church, Charles C and Roy, Ronald A},
  journal={The Journal of the Acoustical Society of America},
  volume={91},
  number={6},
  pages={3166--3183},
  year={1992},
  publisher={Acoustical Society of America}
}

@article{plesset1977bubble,
  title={Bubble dynamics and cavitation},
  author={Plesset, Milton S and Prosperetti, Andrea},
  journal={Annual review of fluid mechanics},
  volume={9},
  number={1},
  pages={145--185},
  year={1977},
  publisher={Annual Reviews 4139 El Camino Way, PO Box 10139, Palo Alto, CA 94303-0139, USA}
}

@article{everest1948acoustical,
  title={Acoustical characteristics of noise produced by snapping shrimp},
  author={Everest, F Alton and Young, Robert W and Johnson, Martin W},
  journal={The Journal of the Acoustical Society of America},
  volume={20},
  number={2},
  pages={137--142},
  year={1948},
  publisher={Acoustical Society of America}
}

@article{tang2019bioinspired,
  title={Bioinspired mechanical device generates plasma in water via cavitation},
  author={Tang, Xin and Staack, David},
  journal={Science advances},
  volume={5},
  number={3},
  pages={eaau7765},
  year={2019},
  publisher={American Association for the Advancement of Science}
}

@article{plesset1949dynamics,
  title={The dynamics of cavitation bubbles},
  author={Plesset, Milton S},
  year={1949}
}

@misc{rayleigh1917pressure,
  title={On the pressure developed in a liquid during the collapse of a spherical cavity: Philosophical Magazine Series 6, 34, 94--98},
  author={Rayleigh, L},
  year={1917}
}

@article{barber1992resolving,
  title={Resolving the picosecond characteristics of synchronous sonoluminescence},
  author={Barber, Bradley P and Hiller, Robert and Arisaka, Katsushi and Fetterman, Harold and Putterman, Seth},
  journal={The Journal of the Acoustical Society of America},
  volume={91},
  number={5},
  pages={3061--3063},
  year={1992},
  publisher={Acoustical Society of America}
}

@article{hiller1992spectrum,
  title={Spectrum of synchronous picosecond sonoluminescence},
  author={Hiller, Robert and Putterman, Seth J and Barber, Bradley P},
  journal={Physical review letters},
  volume={69},
  number={8},
  pages={1182},
  year={1992},
  publisher={APS}
}

@misc{jackson1999classical,
  title={Classical electrodynamics},
  author={Jackson, John David},
  year={1999},
  publisher={American Association of Physics Teachers}
}

@article{prosperetti1999old,
  title={Old-fashioned bubble dynamics},
  author={Prosperetti, Andrea},
  journal={Sonochemistry and sonoluminescence},
  pages={39--62},
  year={1999},
  publisher={Springer}
}

@article{prosperetti1986bubble,
  title={Bubble dynamics in a compressible liquid. Part 1. First-order theory},
  author={Prosperetti, A and Lezzi, A},
  journal={Journal of Fluid Mechanics},
  volume={168},
  pages={457--478},
  year={1986},
  publisher={Cambridge University Press}
}

@book{brennen2014cavitation,
  title={Cavitation and bubble dynamics},
  author={Brennen, Christopher E},
  year={2014},
  publisher={Cambridge University Press}
}

@book{yasui2018acoustic,
  title={Acoustic cavitation and bubble dynamics},
  author={Yasui, Kyuichi},
  year={2018},
  publisher={Springer}
}

@article{suslick2008inside,
  title={Inside a collapsing bubble: sonoluminescence and the conditions during cavitation},
  author={Suslick, Kenneth S and Flannigan, David J},
  journal={Annu. Rev. Phys. Chem.},
  volume={59},
  pages={659--683},
  year={2008},
  publisher={Annual Reviews}
}

@article{leighton2007derivation,
  title={Derivation of the Rayleigh-Plesset equation in terms of volume},
  author={Leighton, TG},
  year={2007},
  publisher={Institute of Sound and Vibration Research, University of Southampton}
}

@article{lentz1995mie,
  title={Mie scattering from a sonoluminescing air bubble in water},
  author={Lentz, WJ and Atchley, Anthony A and Gaitan, D Felipe},
  journal={Applied optics},
  volume={34},
  number={15},
  pages={2648--2654},
  year={1995},
  publisher={Optical Society of America}
}

@article{gompf2000mie,
  title={Mie scattering from a sonoluminescing bubble with high spatial and temporal resolution},
  author={Gompf, B and Pecha, R},
  journal={Physical Review E},
  volume={61},
  number={5},
  pages={5253},
  year={2000},
  publisher={APS}
}

@article{lezzi1987bubble,
  title={Bubble dynamics in a compressible liquid. Part 2. Second-order theory},
  author={Lezzi, A and Prosperetti, A},
  journal={Journal of Fluid Mechanics},
  volume={185},
  pages={289--321},
  year={1987},
  publisher={Cambridge University Press}
}

@article{prosperetti1988nonlinear,
  title={Nonlinear bubble dynamics},
  author={Prosperetti, Andrea and Crum, Lawrence A and Commander, Kerry W},
  journal={The Journal of the Acoustical Society of America},
  volume={83},
  number={2},
  pages={502--514},
  year={1988},
  publisher={Acoustical Society of America}
}

@article{keller1956damping,
  title={Damping of underwater explosion bubble oscillations},
  author={Keller, Joseph B and Kolodner, Ignace I},
  journal={Journal of applied physics},
  volume={27},
  number={10},
  pages={1152--1161},
  year={1956},
  publisher={American Institute of Physics}
}

@article{lofstedt1995sonoluminescing,
  title={Sonoluminescing bubbles and mass diffusion},
  author={L{\"o}fstedt, Ritva and Weninger, Keith and Putterman, Seth and Barber, Bradley P},
  journal={Physical Review E},
  volume={51},
  number={5},
  pages={4400},
  year={1995},
  publisher={APS}
}

@article{barber1997defining,
  title={Defining the unknowns of sonoluminescence},
  author={Barber, Bradley P and Hiller, Robert A and L{\"o}fstedt, Ritva and Putterman, Seth J and Weninger, Keith R},
  journal={Physics Reports},
  volume={281},
  number={2},
  pages={65--143},
  year={1997},
  publisher={Elsevier}
}

@article{keller1980bubble,
  title={Bubble oscillations of large amplitude},
  author={Keller, Joseph B and Miksis, Michael},
  journal={The Journal of the Acoustical Society of America},
  volume={68},
  number={2},
  pages={628--633},
  year={1980},
  publisher={Acoustical Society of America}
}

@incollection{yasui2015dynamics,
  title={Dynamics of acoustic bubbles},
  author={Yasui, Kyuichi},
  booktitle={Sonochemistry and the Acoustic Bubble},
  pages={41--83},
  year={2015},
  publisher={Elsevier}
}

@article{matula2000single,
  title={Single-bubble sonoluminescence in microgravity},
  author={Matula, Thomas J},
  journal={Ultrasonics},
  volume={38},
  number={1-8},
  pages={559--565},
  year={2000},
  publisher={Elsevier}
}

@article{matula1999inertial,
  title={Inertial cavitation and single--bubble sonoluminescence},
  author={Matula, Thomas J},
  journal={Philosophical Transactions of the Royal Society of London. Series A: Mathematical, Physical and Engineering Sciences},
  volume={357},
  number={1751},
  pages={225--249},
  year={1999},
  publisher={The Royal Society}
}

@article{flannigan2006measurement,
  title={Measurement of pressure and density inside a single sonoluminescing bubble},
  author={Flannigan, David J and Hopkins, Stephen D and Camara, Carlos G and Putterman, Seth J and Suslick, Kenneth S},
  journal={Physical review letters},
  volume={96},
  number={20},
  pages={204301},
  year={2006},
  publisher={APS}
}

@article{flannigan2005plasma,
  title={Plasma line emission during single-bubble cavitation},
  author={Flannigan, David J and Suslick, Kenneth S},
  journal={Physical review letters},
  volume={95},
  number={4},
  pages={044301},
  year={2005},
  publisher={APS}
}

@article{weninger1995sonoluminescence,
  title={Sonoluminescence from single bubbles in nonaqueous liquids: new parameter space for sonochemistry},
  author={Weninger, Keith and Hiller, Robert and Barber, Bradley P and Lacoste, David and Putterman, Seth J},
  journal={The Journal of Physical Chemistry},
  volume={99},
  number={39},
  pages={14195--14197},
  year={1995},
  publisher={ACS Publications}
}

@article{hiller1994effect,
  title={Effect of noble gas doping in single-bubble sonoluminescence},
  author={Hiller, Robert and Weninger, Keith and Putterman, Seth J and Barber, Bradley P},
  journal={Science},
  volume={266},
  number={5183},
  pages={248--250},
  year={1994},
  publisher={American Association for the Advancement of Science}
}

@article{lofstedt1993toward,
  title={Toward a hydrodynamic theory of sonoluminescence},
  author={L{\"o}fstedt, Ritva and Barber, Bradley P and Putterman, Seth J},
  journal={Physics of Fluids A: Fluid Dynamics},
  volume={5},
  number={11},
  pages={2911--2928},
  year={1993},
  publisher={American Institute of Physics}
}

@article{an2006mechanism,
  title={Mechanism of single-bubble sonoluminescence},
  author={An, Yu},
  journal={Physical Review E},
  volume={74},
  number={2},
  pages={026304},
  year={2006},
  publisher={APS}
}

@article{an2008spectral,
  title={Spectral lines of OH radicals and Na atoms in sonoluminescence},
  author={An, Yu and Li, Chaohui},
  journal={Physical Review E},
  volume={78},
  number={4},
  pages={046313},
  year={2008},
  publisher={APS}
}

@article{an2009diagnosing,
  title={Diagnosing temperature change inside sonoluminescing bubbles by calculating line spectra},
  author={An, Yu and Li, Chaohui},
  journal={Physical Review E},
  volume={80},
  number={4},
  pages={046320},
  year={2009},
  publisher={APS}
}

@article{lohse1997sonoluminescing,
  title={Sonoluminescing air bubbles rectify argon},
  author={Lohse, Detlef and Brenner, Michael P and Dupont, Todd F and Hilgenfeldt, Sascha and Johnston, Blaine},
  journal={Physical review letters},
  volume={78},
  number={7},
  pages={1359},
  year={1997},
  publisher={APS}
}

@article{sivasubramanian2002temperature,
  title={Temperature of a Compressed Bubble with Application to Sonoluminescence},
  author={Sivasubramanian, S and Widom, A and Srivastava, YN},
  journal={arXiv preprint cond-mat/0207465},
  year={2002}
}

@article{hilgenfeldt1999simple,
  title={A simple explanation of light emission in sonoluminescence},
  author={Hilgenfeldt, Sascha and Grossmann, Siegfried and Lohse, Detlef},
  journal={Nature},
  volume={398},
  number={6726},
  pages={402--405},
  year={1999},
  publisher={Nature Publishing Group}
}

@article{hilgenfeldt1999sonoluminescence,
  title={Sonoluminescence light emission},
  author={Hilgenfeldt, Sascha and Grossmann, Siegfried and Lohse, Detlef},
  journal={Physics of fluids},
  volume={11},
  number={6},
  pages={1318--1330},
  year={1999},
  publisher={American Institute of Physics}
}

@article{yasui1999mechanism,
  title={Mechanism of single-bubble sonoluminescence},
  author={Yasui, Kyuichi},
  journal={Physical review E},
  volume={60},
  number={2},
  pages={1754},
  year={1999},
  publisher={APS}
}

@article{geltman1973free,
  title={Free-free radiation in electron-neutral atom collisions},
  author={Geltman, Sydney},
  journal={Journal of Quantitative Spectroscopy and Radiative Transfer},
  volume={13},
  number={7},
  pages={601--613},
  year={1973},
  publisher={Elsevier}
}

@misc{schroeder1999introduction,
  title={An introduction to thermal physics},
  author={Schroeder, Daniel V},
  year={1999},
  publisher={American Association of Physics Teachers}
}

@book{zel2002physics,
  title={Physics of shock waves and high-temperature hydrodynamic phenomena},
  author={Zel'Dovich, Ya B and Raizer, Yu P},
  year={2002},
  publisher={Courier Corporation}
}

@article{taylor1969experimental,
  title={Experimental determination of the cross-sections for neutral Bremsstrahlung: I. Ne, Ar and Xe},
  author={Taylor, Raymond L and Caledonia, George},
  journal={Journal of Quantitative Spectroscopy and Radiative Transfer},
  volume={9},
  number={5},
  pages={657--679},
  year={1969},
  publisher={Elsevier}
}

@article{eckert2015aether,
  title={From aether impulse to QED: Sommerfeld and the Bremsstrahlen theory},
  author={Eckert, Michael},
  journal={Studies in History and Philosophy of Science Part B: Studies in History and Philosophy of Modern Physics},
  volume={51},
  pages={9--22},
  year={2015},
  publisher={Elsevier}
}

@misc{griffiths2005introduction,
  title={Introduction to electrodynamics},
  author={Griffiths, David J},
  year={2005},
  publisher={American Association of Physics Teachers}
}

@article{hammer2002light,
  title={Light emission of sonoluminescent bubbles containing a rare gas and water vapor},
  author={Hammer, Dominik and Frommhold, Lothar},
  journal={Physical Review E},
  volume={65},
  number={4},
  pages={046309},
  year={2002},
  publisher={APS}
}

@article{frommhold1998electron,
  title={Electron-atom bremsstrahlung and the sonoluminescence of rare gas bubbles},
  author={Frommhold, Lothar},
  journal={Physical Review E},
  volume={58},
  number={2},
  pages={1899},
  year={1998},
  publisher={APS}
}

@article{yasui2001effect,
  title={Effect of liquid temperature on sonoluminescence},
  author={Yasui, Kyuichi},
  journal={Physical Review E},
  volume={64},
  number={1},
  pages={016310},
  year={2001},
  publisher={APS}
}

@article{unnikrishnan1996comment,
  title={Comment on “Sonoluminescence as Quantum Vacuum Radiation”},
  author={Unnikrishnan, CS and Mukhopadhyay, Shomeek},
  journal={Physical review letters},
  volume={77},
  number={22},
  pages={4690},
  year={1996},
  publisher={APS}
}

@article{lambrecht1997comment,
  title={Comment on" Sonoluminescence as Quantum Vacuum Radiation''},
  author={Lambrecht, Astrid and Jaekel, Marc-Thierry and Reynaud, Serge},
  journal={Physical review letters},
  volume={78},
  pages={2267},
  year={1997}
}

@article{garcia1997comment,
  title={Comment on “Sonoluminescence as Quantum Vacuum Radiation”},
  author={Garcia, N and Levanyuk, AP},
  journal={Physical review letters},
  volume={78},
  number={11},
  pages={2268},
  year={1997},
  publisher={APS}
}

@article{milton1996comment,
  title={Comment on" Sonoluminescence as Quantum Vacuum Radiation"},
  author={Milton, Kimball A},
  journal={arXiv preprint quant-ph/9608003},
  year={1996}
}

\end{document}